# A photoconductor intrinsically has no gain

Yaping Dan[1]*, Xingyan Zhao[1], Kaixiang Chen[1] and Abdelmadjid Mesli[2]

[1]University of Michigan – Shanghai Jiao Tong University Joint Institute, Shanghai Jiao Tong University, 800 Dong Chuan Road, Shanghai, China

[2]Institute Matériaux Microélectronique Nanosciences de Provence, UMR 6242 CNRS, Université Aix-Marseille, 13397 Marseille Cedex 20, France

*Correspondence should be addressed to: yaping.dan@sjtu.edu.cn

Abstract

In the past 50 years, the high gain in quantum efficiency of photoconductors is often explained by a widely accepted theory in which the photogain is proportional to the minority carrier lifetime and inversely proportional to the carrier transit time across the photoconductor. It occasionally misleads scientists to believe that a high-speed and high-gain photodetector can be made simply by shortening the device length. The theory is derived on the assumption that the distribution of photogenerated excess carriers is spatially uniform. In this Letter, we find that this assumption is not valid for a photoconductive semiconductor due to the metal-semiconductor boundary at the two metal electrodes inducing carrier confinement. By solving the continuity equation and performing numerical simulations, we conclude that a photoconductor intrinsically has no gain or at least no high gain, no matter how short the transit time and how long the minority lifetime is. The high gain observed in experiments comes from other extrinsic effects such as defects, surface states and surface depletion regions that localize excess minority carriers, leaving a large number of excess majority carriers accumulated in the conduction channel for the photogain. Following the Ohm's Law, a universal equation governing the photogain in a photoconductor is established at the end of this Letter.

It is well known that avalanche photodiodes and bipolar phototransistors have gain in terms of quantum efficiency. A photoconductive semiconductor having gain is surprising, but it is clearly written in the classical semiconductor physics textbooks[1-3] and widely accepted by the research community for decades[4-8]. The gain theory was derived in 1950s[9], which concluded that the gain of a photoconductor is equal to the recombination lifetime of minority carriers divided by the transit time that the carriers take to transport across the semiconductor between the two contacts of the device. Therefore, a photoconductor will intrinsically have a large gain if the transit time is much shorter than the recombination lifetime. The physical explanation for the gain is that the short transit time allows the photogenerated carriers to circulate in the circuit multiple times before recombination, equivalent to generating many times more photoexcited carriers[1]. We call this gain theory as "recycling gain mechanism" for convenience.

Conceptually, according to the theory, the recycling of charge carriers increases the number of collected carriers but not the concentration of excess carriers in the device. The theory will inevitably lead to the conclusion of no gain in photoconductivity, which however is in contradiction with most of the experimental observations[4, 8]. Quantitatively, there is a huge disparity between the gains predicted by the theory and those measured in experiments. For instance, Matsuo *et al* [10] observed in 1984 that the gain of GaAs photoconductive detectors predicted by the recycling gain theory is 3 to 4 orders of magnitude smaller than the gain measured in the experiments. Similar observations have been made persistently by other researchers in the past several decades[4, 11, 12]. Some argued that this disparity is due to the carrier trapping by surface trap states or charge separation by built-in electric fields that increases the recombination lifetime of minority carriers [6, 13, 14]. Others even mixed up the concepts of trap-emission and minority recombination lifetimes[4], using the long trap lifetime to replace the short minority recombination lifetime to explain away the disparity. Up to date, this gain theory is still being widely used to explain the observed photoconductive gain in photoconductors based on quantum dots [15], nanowires [7, 16] and more recently 2-dimensional materials[17, 18].

In this Letter, we find that this well-known recycling gain theory is highly questionable because its derivation does not consider the metal-semiconductor boundary and is based on the assumption that the concentration of photogenerated excess carriers in the photoconductor is uniformly distributed. However, for a semiconductor in contact with metal (as electrodes), the photogenerated excess carriers in the semiconductor are always spatially non-uniform due to the carrier confinement by the metal-semiconductor boundary, and therefore electric field dependent. By solving the continuity equation and performing simulations using the commercial device simulator, we conclude that a photoconductor intrinsically has no gain or at least no high gain. It means that, for a photoconductive semiconductor in contact with metal electrodes, the theoretical gain will never be greater than 1 or at most not greater than the ratio of the majority to minority mobility, no matter how short the transit time and how long the minority lifetime is. In the latter case, it might be higher than 1 if the majority carrier mobility is larger than the minority carrier mobility. The high gain observed in experiments comes from other extrinsic effects such as the trapping effect of defects, surface states and/or surface depletion regions that will localize excess minority carriers and leave a large number of excess majority counterparts accumulated in the conduction channel, leading thus to the observed high photogain. Following the Ohm's Law, a new equation governing the photogain in photoconductors is established at the end of this Letter.

Let us first go through the theoretical derivation of the recycling gain mechanism in the classical semiconductor physics textbooks[1]. The gain G of a photoconductor following the definition of internal quantum efficiency is defined as the number of photogenerated charge carriers collected by the electrodes divided by the number of photons absorbed in the semiconducting photoconductor.

$$G = \frac{N_{electron}}{N_{photon}} = \frac{J_{ph} \cdot A_c/e}{P_{abs}/\hbar\omega} \qquad eq.(1)$$

where $J_{ph}$ is the photocurrent density, $A_c$ the cross-sectional area of the device, $e$ the charge unit, and $\hbar\omega$ the photon energy. The denominator $P_{abs}/\hbar\omega$ is the total number of photons absorbed per second in the device. If we assume one absorbed photon generating one electron-hole pair, the carrier generation rate is equal to $g = P_{abs}/\hbar\omega /V$ where $V$ is the device volume given by $V = A_c \times L$ with $L$ being the length between the two electrodes of the photoconductor. Then eq.(1) can be further written as *eq.(2)*

$$G = \frac{N_{electron}}{N_{photon}} = \frac{J_{ph} \cdot A_c/e}{P_{abs}/\hbar\omega} = \frac{J_{ph}}{egL} \qquad eq.(2)$$

The photocurrent density equals to $J_{ph} = e(\mu_n \Delta n + \mu_p \Delta p) \cdot E$ in which $E$ is the electric field intensity, $\mu_n$ and $\mu_p$ are the electron and hole mobility, and $\Delta n$ and $\Delta p$ are the photogenerated electron and hole concentration, respectively, and $\Delta n = \Delta p$ as the excess carriers are generated in pairs. In general, the photogenerated minority carrier concentration can be written as *eq.(3)*

$$\Delta n = g \cdot \tau_n \qquad eq.(3)$$

where $g$ is the generation rate and $\tau_n$ is the recombination lifetime of minority electrons in a semiconductor. Note that $\tau_n$ is determined by the quality of the semiconductor at the atomic level. The incorporation of a given amount of defects and impurities is unavoidable.

By plugging *eq.*(3) into *eq.*(2), we get:

$$G = \frac{\tau_n(\mu_n+\mu_p)E}{L} = \frac{\tau_n}{\tau_t}\left(1 + \frac{\mu_p}{\mu_n}\right) \qquad eq.(4)$$

in which $\tau_t = L/(\mu_n E)$ is the transit time for the minority electrons to transport between the two contact electrodes of the photoconductor. Note that the transit time has a low limit due to the velocity saturation. Nevertheless the gain according to *eq.*(4) can still be very high if the recombination lifetime is much longer than the transit time by applying a large electric voltage on a short device.

*eq.*(4) is the theoretical foundation of the recycling gain mechanism. The main problem of this theory originates from *eq.*(3) which is assumed to be spatially uniform and independent of the electric field intensity. But for a semiconductor in contact with metal, the distribution of photogenerated excess carriers is always non-uniform and therefore readily skewed by the electric field, resulting in the voltage-dependent excess carrier concentration. This can be seen clearly from the continuity equation. For a uniformly doped p-type semiconductor under small injection condition, the continuity equation at steady state for minority carriers is expressed as:

$$D_n \frac{\partial^2 \Delta n}{\partial x^2} + \mu_n E \frac{\partial \Delta n}{\partial x} + \mu_n \Delta n \frac{\partial E}{\partial x} - \frac{\Delta n}{\tau_n} + g_n = 0 \qquad eq.(5)$$

where $D_n, \mu_n, \tau_n$ and $g_n$ are the diffusion constant, mobility, recombination lifetime and generation rate of minority electrons, respectively. With voltage bias, the electric field inside the device may be uniform (Ohmic contact) but always nonzero. The uniform electric field will zero out the third term. On the other hand, the second and consequently the first term are zero only if the excess carriers are uniformly distributed. In this case, *eq.*(3) is valid.

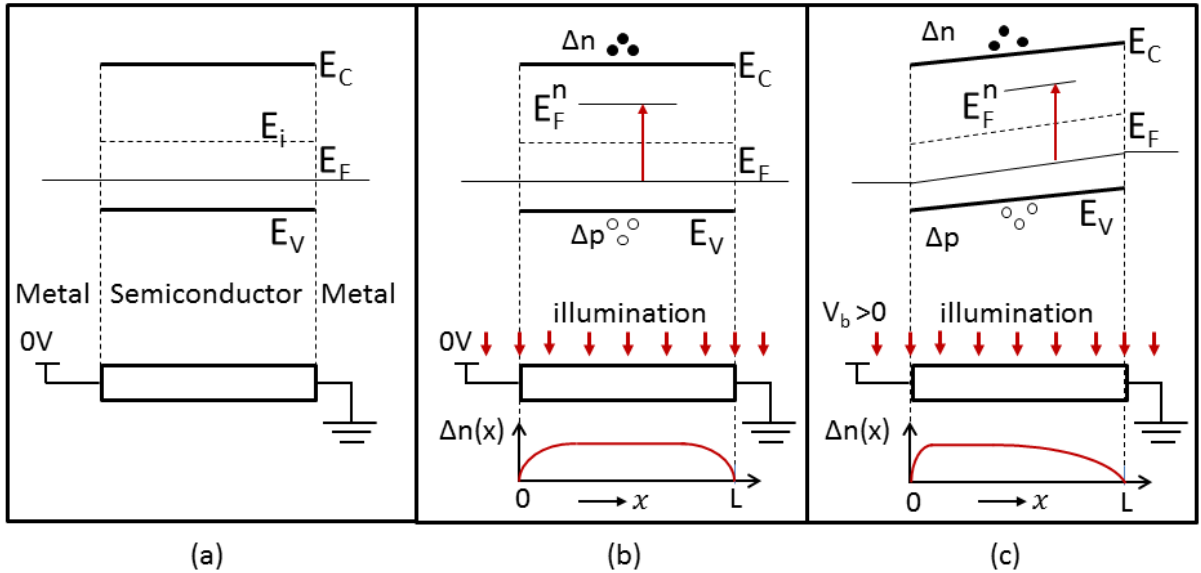

Figure 1. Energy band diagram of a photoconductor (a) in dark, (b) under light illumination with zero voltage bias and (c) under light illumination with nonzero voltage bias. The semiconductor is assumed to have the same work function with the metal.

However, the distribution of photogenerated excess carriers in a semiconductor in contact with metal is always non-uniform, as shown in Fig.1. For simplicity, we assume that the semiconductor is in Ohmic contacts with the metal electrodes by having the same work function with the metal. (Other types of Ohmic contacts formed by silicides or tunneling Schottky barriers are also studied. See SI Section 1. The results are consistent with what we find below.) There is no energy band bending when they are in contact (Fig.1a). Light is uniformly illuminated on the device from the vertical direction, as shown in the sketch of Fig.1b and c. Excess minority electrons are excited in the conduction band in the semiconductor and no excess electrons will be generated in the metal. To maintain continuity, the concentration of excess minority electrons has to be zero at the semiconductor-metal interface, resulting in excess electrons in the semiconductor diffusing towards the metal, as shown in the bottom sketch of Fig.1b. At zero voltage bias, the electron diffusion is anti-symmetric with no net photocurrent flow in the circuit. At non-zero bias, the electric field will skew the anti-symmetric transport of excess electrons (Fig.1c), creating a net photocurrent. Clearly, the concentration of photogenerated excess carriers is spatially voltage-dependent instead of following the simple expression of *eq.*(3). If *eq.*(3) cannot hold, then the gain expression *eq.*(4) derived on the basis of *eq.*(3) is questionable.

To derive the correct expression for the gain, we need to first find the minority carrier distribution by solving the continuity equation *eq.*(5) with the assumption of uniform electric field (the third term is zero). This assumption is valid for a uniformly doped semiconductor with Ohmic contact at small injection condition. By applying the boundary conditions $\Delta n = 0$ at both $x = 0$ and $x = L$, we find:

$$\Delta n(x) = g_n \tau_n \frac{1-\exp(\lambda_2 L)}{\exp(\lambda_2 L)-\exp(\lambda_1 L)} \cdot \exp(\lambda_1 x) - g_n \tau_n \frac{1-\exp(\lambda_1 L)}{\exp(\lambda_2 L)-\exp(\lambda_1 L)} \cdot \exp(\lambda_2 x) + g_n \tau_n \qquad eq.(6)$$

where $\lambda_{1,2} = \frac{-L_{dr}(E) \pm \sqrt{L_{dr}^2(E)+4L_D^2}}{2L_D^2}$ ("+" for $\lambda_1$ and "-" for $\lambda_2$) with the drift length $L_{dr} = \mu_n \tau_n E$ and the diffusion length $L_D = \sqrt{D_n \tau_n}$.

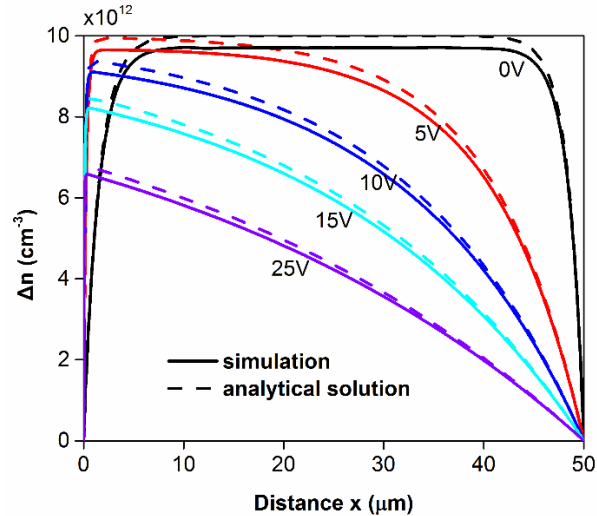

Figure 2. Spatial distribution of photogenerated excess minority carriers in a photoconductor. Dotted lines are the solutions of the continuity equation and solid lines are the simulation results. Velocity saturation is excluded from the simulation.

To validate the solution of the continuity equation given by *eq.*(6), we performed numerical simulations on a silicon photoconductor using the DEVICE module of the commercial software Lumerical. The software module numerically solves the Poisson's equation and the continuity equations for minority and majority carriers. It can catch the transport behavior of both types of carriers, providing more realistic results. We suppose that the device under simulation is 50 μm long and 1 μm × 1 μm in cross-section. The p-type doping concentration is $10^{17}$ cm$^{-3}$ and the generation rate is spatially uniform at $10^{22}$ cm$^{-3}$/s for the sake of clarity. The recombination lifetime of minority carriers is set at 1 ns due to, for instance, defects and impurities as recombination centers. The mobility for electrons and holes is 875 cm$^2$/Vs and 337 cm$^2$/Vs[2], respectively. Different lifetime and mobility for electrons and holes will not change the conclusion (see more discussions later). As we show later, the photocurrent will saturate at high voltage bias. To illustrate that the photocurrent saturation is not caused by the velocity saturation, the velocity saturation effect is excluded from the simulation. The spatial distributions of photogenerated excess carriers are plotted in Fig.2. The solid and dotted lines denote the concentrations of excess minority carriers $\Delta n$ given by the device simulator and the equation $eq.(6)$, respectively. It is clear that the simulation results and the solutions of the continuity equation are almost identical.

$$J_n = -eD_n g_n \tau_n \frac{\lambda_2(1-e^{\lambda_1 L})(1+e^{\lambda_2 L})-\lambda_1(1-e^{\lambda_2 L})(1+e^{\lambda_1 L})}{e^{\lambda_2 L}-e^{\lambda_1 L}} = \begin{cases} eg_n \tau_n \mu_n E & E \to 0, L_D \ll L \\ eg_n L & E \to \infty \end{cases} \quad eq.(7)$$

As expected, the anti-symmetric distribution of the excess carriers is skewed by the electric field (Fig.2), which will create net photocurrent in the circuit. The equation for the minority photocurrent is given by $eq.(7)$. This expression is rather complicated but it can be simplified to the forms that we are more familiar with at two extreme cases. The first case is when the electric field intensity is close to zero. The drift length ($L_{dr}$) is then nearly zero, much smaller than the diffusion length ($L_D$). Logically, the transit time of minority carriers will be significantly longer than the recombination lifetime, i.e. $\tau_n/\tau_t \ll 1$. In this case, the spatial distribution of photogenerated excess minority carriers remains almost anti-symmetric. If the diffusion length $L_D$ is much smaller than the device length L, then the excess minority carriers distribute almost uniformly in the semiconductor. The uniform distribution of photogenerated carriers and electric field will zero out the first three terms in eq.(5), resulting in $\Delta n = g_n \tau_n$. The photocurrent density of minority excess carriers will then be given by $J_n = eg_n \tau_n \mu_n E$, consistent with the simplification in $eq.(7)$ for a small electric field. For the case that the electric field intensity E approaches very large values, the excess carrier distribution is strongly skewed (like the curve at 25V bias in Fig. 2). The equation $\Delta n = g_n \tau_n$ will never satisfy. In this case, the transit time will be much shorter than the recombination lifetime, i.e. $\tau_n/\tau_t \gg 1$. The minority electron photocurrent density saturates to $J_n = eg_n L$ instead of linearly going up, as shown in $eq.(7)$. This is not surprising if we take into account the fact that the concentration of excess minority carriers decreases as the bias increases, as shown in Fig.2.

The photogenerated excess majority carriers $\Delta p$ also contribute to the photocurrent. Note that the semiconductor is doped. There is a large background dark current contributed by the majority carriers. The continuity equation for majority carriers is a nonlinear differential equation, from which it is difficult to analytically solve the spatial distribution of the excess majority carriers. Nevertheless it is known (we also verified by simulations, see SI Section 2) that the spatial distributions of excess majority and minority carriers are nearly identical if the external electric field is not too high, regardless of the difference in mobility for minority and majority carriers. This phenomenon is called ambipolar transport[1]. As stated above, the excess minority electrons $\Delta n = g_n \tau_n$ are mostly uniformly distributed and the electron photocurrent density is given by $J_n = e g_n \tau_n \mu_n E$ on the condition that the electric field intensity E is not strong and the diffusion length is much smaller than the device length L (Fig.1b). Due to the ambipolar transport phenomenon, the same conclusion can be reached for the excess majority holes, i.e. $\Delta p = g_n \tau_n$ and $J_p = e g_n \tau_n \mu_p E$. Therefore the total photocurrent density is governed by $J_{ph} = e g_n \tau_n (\mu_n + \mu_p) E$, which is consistent with the common knowledge and the simulation results shown in Fig.3a at small voltages. In this case, the gain expression given by *eq.*(4) still holds except that the gain is much smaller than 1, because $\tau_n/\tau_t \ll 1$ at small electric field intensity as previously analyzed for the minority carriers.

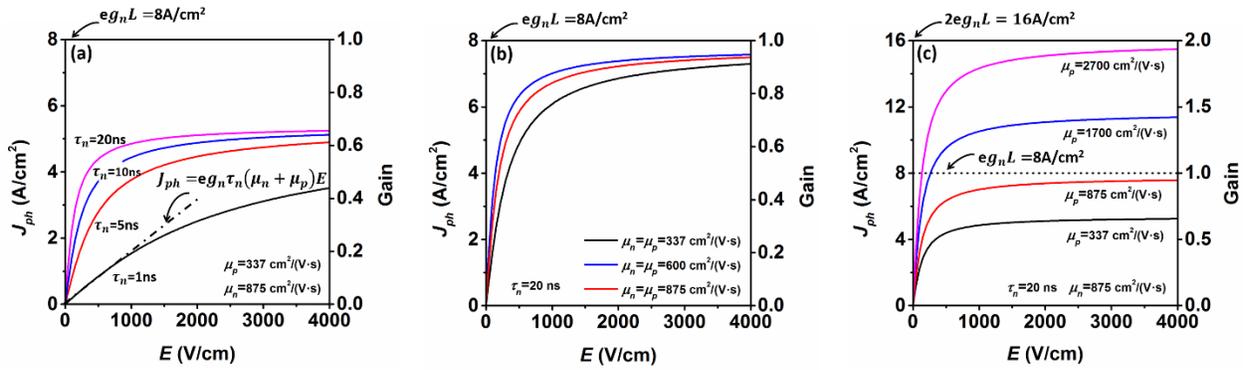

Figure 3. Photocurrent density vs electric field intensity. The device has the same parameters with the one in Fig.2 except for the minority recombination lifetime and mobility. (a) Photocurrent saturates to a value smaller than $egL$ if the majority carrier mobility is smaller than the minority carrier mobility. A longer minority carrier lifetime will not make the photocurrent saturate to a higher value but saturate faster. (b) Photocurrent saturates to $egL$ when the electron and hole mobility are equal. (c) Photocurrent saturates to a value larger than $egL$ when the majority mobility is bigger than the minority mobility.

At high electric field, the total photocurrent density, instead of increasing linearly, saturates to a value that is only a fraction of $J_{ph} = e g_n L$ (gain less than 1, see *eq.*(2)), although the excess electron current approaches $J_n = e g_n L$ (eq.(6)). This is because the slower excess majority holes will accumulate in the semiconductor (the majority hole mobility is smaller than the minority electron mobility in Fig.3a), inducing a small built-in electric field that partially cancels out the electron photocurrent. A longer minority lifetime does not increase the total photocurrent. Instead, it will only make the photocurrent saturate at smaller electric field (Fig.3a). If the majority and minority excess carriers have the same mobility, the built-in electric field disappears and the total current saturates at $J_n = e g_n L$ (Fig.3b with gain approaching but never exceeding 1). If the majority carriers have a higher mobility than the minority carriers (Fig.3c), the saturation photocurrent becomes higher than $J_{ph} = eg_n L$, creating some photo gain in the device. As the majority carrier mobility continues to increase, this photogain becomes even higher (Fig.3c), but not higher than the ratio of the majority to minority mobility. Clearly, this small gain cannot explain the widely observed high photo gains in literature. We can therefore conclude that a photoconductor intrinsically has no gain, or at least no high gain, no matter how short the transit time and how long the minority lifetime is.

If a photoconductor intrinsically has no gain or no high gain, then where are the high gains observed in the experiments coming from?

We recently revealed by photo Hall effect measurements[21] that a silicon nanowire with a high photogain always has an unusually high concentration of excess majority carriers compared to minority excess carriers, i.e. $\Delta p \gg \Delta n$ (assume holes as majority carriers). This experimental observation indicates that the assumption of an equal concentration for excess minority and majority carriers ($\Delta p = \Delta n$) may not always be valid during the derivation of *eq*.(4). To make it more general, we relax this constraint by defining $G' = \Delta p / \Delta n$ as a gain in excess carrier concentration so that *eq*.(4) can be rewritten as eq.(8). If $G'$ is large enough, $G$ will be greater than 1 even at small bias voltage.

$$G = \frac{N_{electron}}{N_{photon}} = \frac{J_{ph}}{egL} = \frac{\tau_n}{\tau_t}\left(1 + G'\frac{\mu_p}{\mu_n}\right) \qquad \text{eq. (8)}$$

Now, the question is how it is possible to have excess majority carriers orders of magnitude higher than excess minorities in a photoconductor, since electrons and holes are generated in pairs by light illumination. Indeed, the total number of excess electrons and holes in the photoconductor are always the same. But the excess electrons and holes that contribute to the photoconductivity may not necessarily be the same, because one type of excess carriers (often minorities) may become localized via the trapping effect of defects, surface states and/or surface depletion region. The same number of the other type of counterparts (often majority carriers) is then left in the conductance channel, resulting in an unusually high $G'$ and photogain $G$.

To verify this hypothesis, we performed numerical simulations on a 400nm-thick and 9 µm long silicon slab using the commercial device simulator Silvaco, as shown in Fig.4a. The light is launched perpendicularly from the top and a uniform generation g is assumed in the whole slab. We tuned the concentration of fixed charges and surface states on the top and bottom surface (see SI Section 3). The photogain G in quantum efficiency and $G'$ in terms of gain in excess carrier concentration are plotted in Fig.4b and c, respectively. The photogain G is calculated according to eq.(2). To plot $G' = \Delta p/\Delta n$, we need to find $\Delta p$ and $\Delta n$ in the photoconductor. Due to the confinement at metal-semiconductor contacts and the impact of surface fixed charges and surface states, the spatial distribution of excess electrons and holes are highly non-uniform in the conduction channel, in particular when the concentration of fixed charges and/or surface states is high, resulting in a rather large photogain. It is more appropriate to use the average excess carrier concentration $\Delta p_{avg}$ and $\Delta n_{avg}$ to replace $\Delta p$ and $\Delta n$, respectively. In this case, the effective minority carrier lifetime $\tau_{n,eff}$ can be written as $\tau_{n,eff} = \Delta n_{avg}/g$ where g is the carrier generation rate. When plotting $G/(1 + G'\mu_p/\mu_n)$ versus $\tau_{n,eff}/\tau_t$ with $\tau_t$ being the transit time (Fig.4c), we find that the two terms are equal to each other in a wide range of variation (all data points in Fig. 4b and c), meaning that the correct equation for the photoconductor gain G shall be written as

$$G = \frac{\tau_{n,eff}}{\tau_t}\left(1 + G'\frac{\mu_p}{\mu_n}\right) \qquad \text{eq.(9)}$$

Note that $\tau_{n,eff}$ is not a constant. It depends on many parameters including intrinsic minority carrier lifetime, bias voltage and the density of surface states. If we assume $G'$ equals to 1, the above equation can fit perfectly the nonlinear curves in Fig.3 in which the gain can be found by normalizing the photocurrent density respective to $egL$ (see SI Section 4). It means that the photoconductor will have no gain or at least not a high gain if there is no gain in excess carrier concentration. It is worth pointing out that eq.(9) is actually universal, simply because the photocurrent density equation $J_{ph} = e(\Delta n \cdot \mu_n + \Delta p \cdot \mu_p)E$ is the Ohm's Law and no further assumption is made in the derivation. The classical gain theory contains two mistakes. First, it assumes that the concentration of excess minority carriers is spatially uniform as $\Delta n = g \cdot \tau_n$, which is only true without boundary confinement. Second, it assumes that the excess electrons and holes in the conduction channel are equal in concentration as $\Delta p = \Delta n$, which is only valid in a 'perfect' semiconductor.

In conclusion, the classical gain theory is derived on the severe assumptions that boundary confinements are inexistent, leading to the wrong conclusion that a photoconductor exhibits a high gain when the minority carrier lifetime is long and the carrier transit time is short. This gain theory often misled scientists to believe that high gain and high speed photodetectors can be constructed simply by shortening the device length to minimize the transit time [17]. We prove in this work that a semiconducting photoconductor intrinsically has no gain or at least no high gain in terms of internal quantum efficiency no matter how long the minority carrier lifetime and how short the transit time is. The high photogain observed in experiments originates from a gain in the concentration of excess charge carriers, which is induced by the trapping effect of defects, surface states and/or surface depletion region. Following the Ohm's Law, we derived a universal gain equation for photoconductors, which may guide scientists to design high-performance photodetectors.

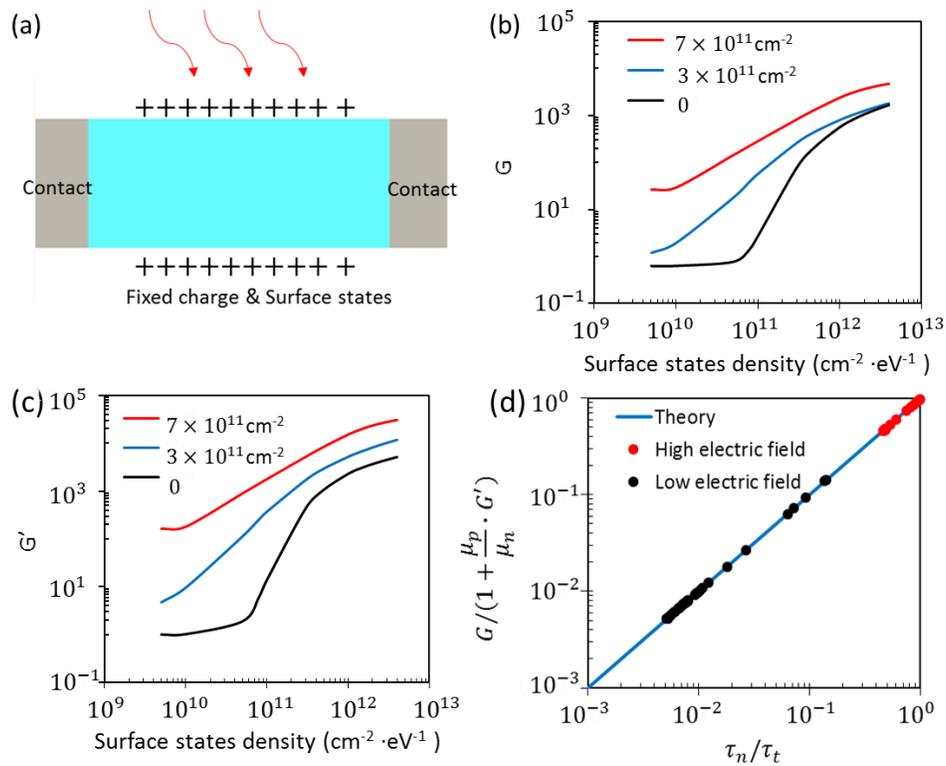

Figure 4. Photogain in presence of fixed charges and surfaces states on the device surfaces. The silicon slab is 400nm thick and p-type with a doping concentration of $1.05\times10^{17}$ cm$^{-3}$. Device schematic is shown in (a). Photogain G in quantum efficiency (b) and gain $G'$ in excess carrier concentration (c) are dependent on surface states density (see SI) and fixed charge concentration of 0, $3\times10^{11}$ cm$^{-2}$ and $7\times10^{11}$ cm$^{-2}$. Eq.(8) is validated by simulations in (d).

**Author Contribution**

Y. D. conceived the concept, derived the theory and wrote the manuscript. X. Z. performed the device simulations of Figure 1-3. K. C. performed the simulations in Fig.4. A. M. commented on the manuscript. All authors reviewed the manuscript.

**Acknowledgement**

The work is supported by the National Science Foundation of China (61376001). We thank Mr. Hongwei Guo and Prof. Yang Xu for useful discussions.

**Conflict of Interests**

The authors declare no conflict of interests.

# A photoconductor intrinsically has no gain

# (Supplementary Information)

Yaping Dan[1]*, Xingyan Zhao[1], Kaixiang Chen[1] and Abdelmadjid Mesli[2]

[1]University of Michigan – Shanghai Jiao Tong University Joint Institute, Shanghai Jiao Tong University, 800 Dong Chuan Road, Shanghai, China

[2]Institute Matériaux Microélectronique Nanosciences de Provence, UMR 6242 CNRS, Université Aix-Marseille, 13397 Marseille Cedex 20, France

*Correspondence should be addressed to: yaping.dan@sjtu.edu.cn

**Section 1. Other forms of Ohmic contacts**

In this section, We performed the 2D simulations using Silvaco on a silicon slab. The following models are included in the simulations: Auger recombination model, SRH recombination model, Universal Schottky tunneling model and Fermi-Dirac carrier statistic model. The generation rate is set uniformly at $2.48\times10^{20}$ cm$^{-3}$. The mobility of electrons and holes are 708 cm$^2$/Vs and 236 cm$^2$/Vs, respectively. The lifetime of electrons and holes are both set at 1ns.

1)  **Ohmic contacts formed by tunneling through a Schottky junction**

The device structure is shown in the panel (a) of Fig.S1. The work function of the metal is 4.8eV and the electron affinity of silicon is 4.17 eV. To facilitate Ohmic contacts by tunneling, the p-type silicon (20nm thick) in contact with the metal is highly doped. The energy band diamgram is shown in the panel (b). The Schottky barrier heigh is about 0.5 eV and the depletion region thickness is about 1-9 nm depending on the doping concentration. When the P$^+$ region is doped at $1\times10^{19}$ cm$^{-3}$, the depletion region thickness is a little large (~9nm). The dark current is nonlinear shown as the red curve in panel (c). When the doping concentration is increased to $3\times 10^{19}$ cm$^{-3}$, the depletion region is narrowed down ~2 nm shown in the closeup plot in panel (b). The dark current becomes linear shown as the black line in panel (c), although at very high voltage it becomes slightly nonlinear (inset in panel (c)). In this case, the photocurrent is nonlinear and saturates to a value lower than eGL as shown in the panel (d), meaning no photogain.

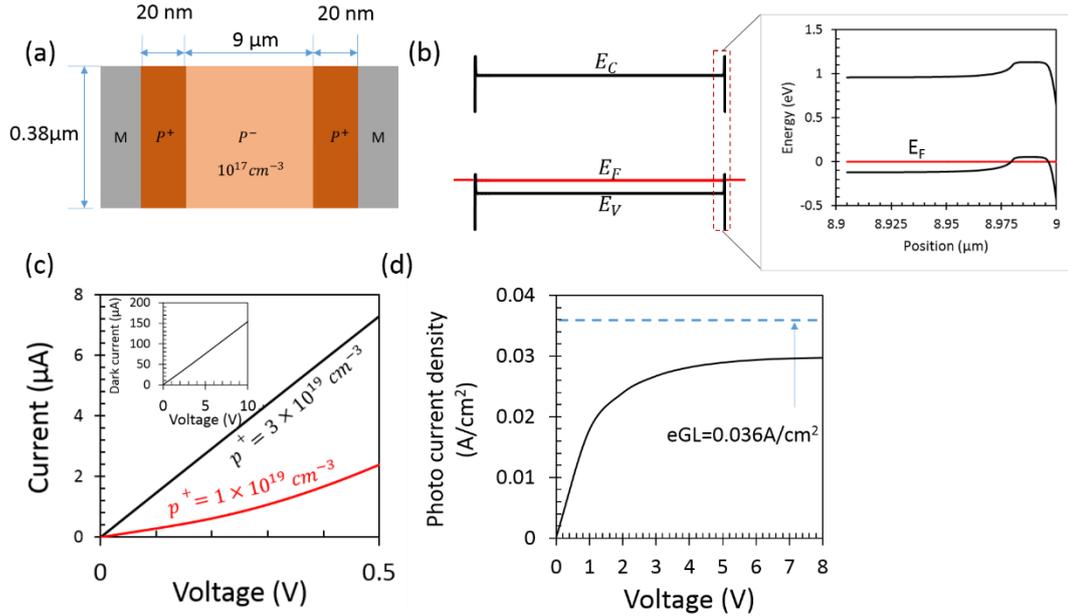

Fig.S1 Schematic, energy band diagram, dark current and photocurrent of a photoconductor in Ohmic contacts with metal by tunneling through a Schottky barrier.

2) **Silicide Ohmic contacts**

We assume that the silicon in contact with the metal is formed a silicide layer after the device is annealed at elavated temperature, as shown in Fig.S2 (a). The silicide is 10nm thick. Silicides are silicon and metal alloys which are semiconductors with a small bandgap. To mimic this case, we assume that the silicide has a bandgap of 0.38eV and the electron affinity larger than silicon, creating an energy band diamgram as shown in Fig. S2 (b). A closeup figure of the band diagram at Si-silicide interface is shown in the inset. Fig.2S (c) shows the dark current vs voltage which is linear. The photocurrent will saturate to a value less than eGL = 0.036 A/cm2 at high voltage bias, as shown in

Fig.S2 (d).

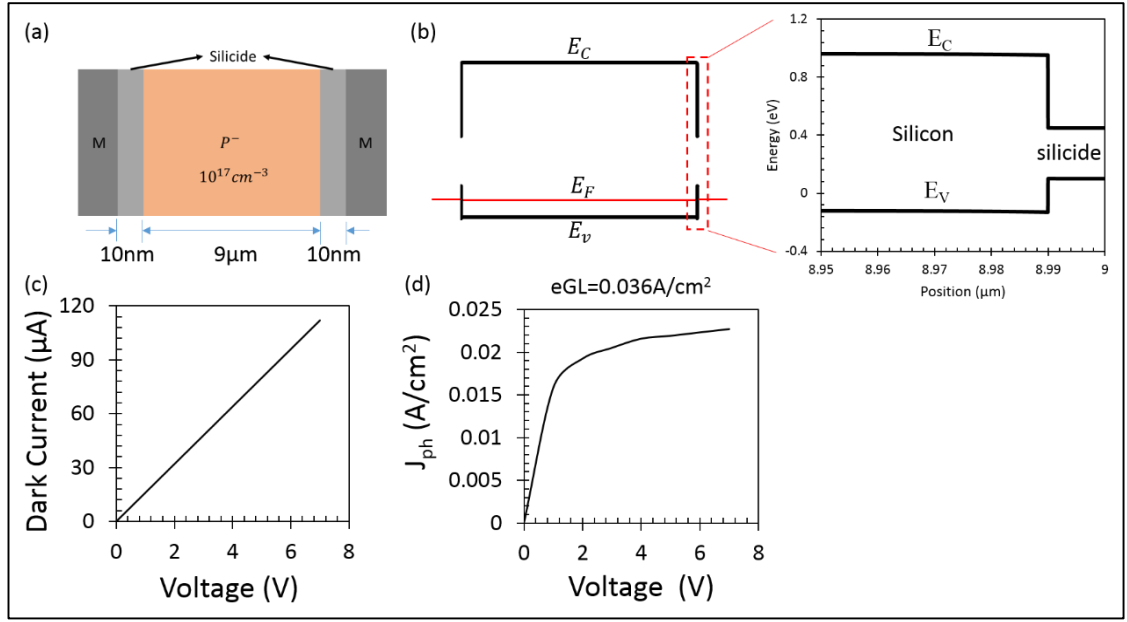

Figure S2. Simulation results for a photoconductor in Ohmic contact with metal by forming silicides at the interface.

**Section 2. Ambipolar transportation**

The simulations were performed by using the Lumerical DEVICE software. The device under simulation is 50 μm long and 1 μm × 1 μm in cross-section. It has a uniform p type doping of $1\times10^{17}$ cm$^{-3}$. The generation rate is a constant of $1\times10^{22}$ cm$^{-3}$s$^{-1}$. Trap-assisted recombination, radiative recombination, surface recombination and auger recombination were excluded from the simulation. Velocity saturation effect is also excluded from the simulation. The mobility and lifetime is varied under different cases. Voltage bias is applied between anode and cathode.

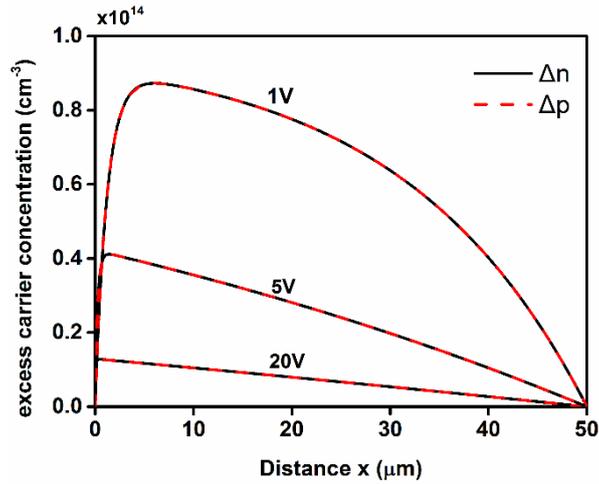

Fig. S3. Spatial distribution of photogenerated excess electrons and holes in a photoconductor under different voltage bias. The electron and hole mobility are 875 cm2/VS and 337 cm2/VS respectively, the minority carrier lifetime is 10ns.

Device simulation shows that the spatial distributions of excess majority and minority carriers are always identical under different voltage bias, showing an ambipolar transport phenomenon.

**Section 3. Surface States and Fixed Charges**

The two dimensional simulations with surface states and fixed charge were performed using Silvaco Atlas software. The fixed charge in the simulations are uniform distributed at the surface, and the density distribution of acceptor-type and donor-type surface states are shown in Fig. S4. During the simulations, the peak of the surface states density varies from $5 \times 10^9$ to $4 \times 10^{12}$ cm$^{-2}$ eV$^{-1}$. The fixed charge density is selt at 0, $3 \times 10^{11}$ and $7 \times 10^{11}$. We assumed that the nanowire length L and thickness w are 9μm and 380nm, respectively. The p-type doping concentration is $1.05 \times 10^{17}$cm$^{-3}$. Apart from that, electron and hole life time are assumed as 200ns and 100ns respectively. The nanowire surfaces are covered with a layer of silicon oxide.

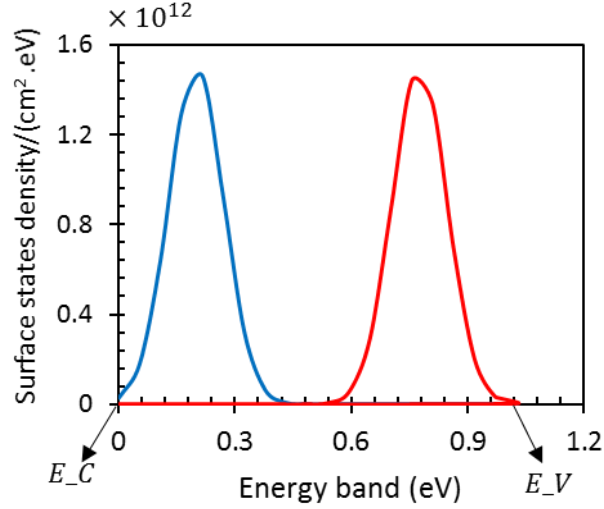

Fig. S4. Density distribution of acceptor-type and donor-type surface states in energy bandgap used in the simulations. Similar distribution patterns of surface states on Si/SiO2 interface have been reported[1].

In the simulations, the hole and electron mobility are set as 236 cm²/Vs and 708 cm²/Vs respectively, and the external voltage U is set as 1V. We can further obtain the transient time for minority carriers $\tau_t$ by

$$\tau_t = \frac{L}{\mu_n \left(\frac{U}{L}\right)}. \tag{S1}$$

The transient time $\tau_t$ can be calculated as $1.14 \times 10^{-9}$s.

By setting the illumination source with wavelength of 460nm and light intensity of 10mW·cm⁻², an average photo generation rate $g$ of $2.58 \times 10^{20}$ cm⁻³s⁻¹ shall be obtained in the simulation. The effective minority carrier life time $\tau_{n,eff}$ can be written as

$$\tau_{n,eff} = \Delta n_{avg}/g \tag{S2}$$

where $\Delta n_{avg}$ is the average excess electron concentration in the nanowire.

Photo gain G expressed by quantum efficiency shall be written as

$$G = \frac{\Delta J}{egL}, \qquad (S3)$$

where $\Delta J$ is the photo current density.

$G'$ is defined as

$$G' = \frac{\Delta p_{avg}}{\Delta n_{avg}}. \qquad (S4)$$

**Section 4. Photo Gain**

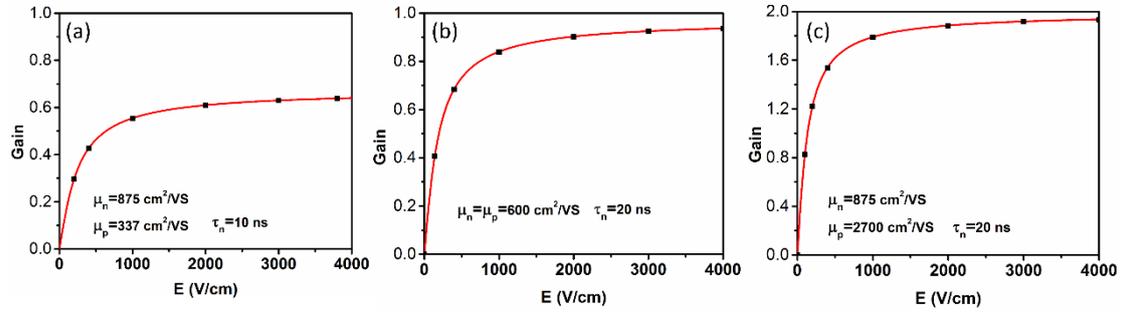

Fig. S5. Simulated photogain (red line) and calculated with eq. (S5) (black square dots).

The red line in Fig.S5 is taken from Fig.3 in the main text. The gain is calculated by divide the photocurrent density with $eg_n L$. The black square dots are calculated as following. The spatial distribution of photogenerated excess electrons $\Delta n(x)$ is obtained from DEVICE simulation, as shown in Figure 2 in the main text. The average excess carrier concentration

$$\Delta n_{avg} = \frac{\int_0^L \Delta n(x)dx}{L}$$

The effective minority carrier lifetime $\tau_{n,eff}$ can be written as $\tau_{n,eff} = \Delta n_{avg}/g$ where g is the carrier generation rate. The gain is then calculated with the following equation:

$$G = \frac{\tau_{n,eff}}{L}(\mu_n + \mu_p)E \qquad \text{eq.(S5)}$$